\documentclass[11pt]{article}
\usepackage{hyperref}
\usepackage{amsmath,amssymb}
\usepackage[active]{srcltx}
\usepackage[all]{xy}
\input diagxy
\usepackage{xcolor}
\font\fr=eufm10 scaled \magstep 1 
\newtheorem{teor}{Theorem}
\newtheorem{prop}{Proposition}

\newtheorem{definition}{Definition}

\newtheorem{remark}{Remark}

\def\beq{\begin{equation}}
\def\eeq{\end{equation}}
\def\bea{\begin{eqnarray}}
\def\eea{\end{eqnarray}}
\def\beann{\begin{eqnarray*}}
\def\eeann{\end{eqnarray*}}
\def\beasn{\begin{sneqnarray}}
\def\eeasn{\end{sneqnarray}}
\def\ben{\begin{enumerate}}
\def\een{\end{enumerate}}
\def\bit{\begin{itemize}}
\def\eit{\end{itemize}}

\def\derpar#1#2{\frac{\partial{#1}}{\partial{#2}}}

\def\qed{\ifvmode\removelastskip\fi
{\unskip\nobreakHfil\penalty50Hbox{}\nobreakHfil
\hbox{\vrule height1.2ex width1.2ex}\parfillskip=0pt
\finalhyphendemerits=0 \par\smallskip}}

\def\vf{\mbox{\fr X}}
\def\df{{\mit\Omega}}
\def\Lag{{\cal L}}
\def\L{{\cal L}}

\def\d{{\rm d}}
\def\H{{\cal H}}

\def\Real{\mathbb{R}}
\def\R{\mathbb{R}}

\newcommand{\Reeb}{\mathcal{R}}
\def\Tan{{\rm T}}
\def\Lie{\mathop{\rm L}\nolimits}
\def\inn{\mathop{i}\nolimits}
\def\Cinfty{{\rm C}^\infty}

\title{UNIFIED LAGRANGIAN-HAMILTONIAN FORMALISM FOR CONTACT SYSTEMS}
\author{\sffamily 
\sc $^a$Manuel de Le\'on,
$^b$Jordi Gaset, 
$^c$Manuel La\'inz, \\
\sc $^d$Xavier Rivas,
$^d$Narciso Rom\'an-Roy%
\thanks{emails: 
mdeleon@icmat.es
jordi.gaset@uab.cat,
manuel.lainz@icmat.es
xavier.rivas@upc.edu,
narciso.roman@upc.edu}
\\[1ex]
\normalsize\itshape\sffamily 
$^a$Instituto de Ciencias Matem\'aticas,
Consejo Superior de Investigaciones Cient\'ificas\\
\normalsize\itshape\sffamily 
and Real Academia de Ciencias, Madrid, Spain.
\\[0.1ex]
\normalsize\itshape\sffamily 
$^b$Department of Physics,
Universitat Aut\`onoma de Barcelona,
Bellaterra, Spain.
\\[1ex]
\normalsize\itshape\sffamily 
$^c$Instituto de Ciencias Matem\'aticas,
Consejo Superior de Investigaciones Cient\'ificas, Madrid, Spain.
\\[0.1ex]
\normalsize\itshape\sffamily 
$^d$Department of Mathematics,
Universitat Polit\`ecnica de Catalunya,
Barcelona, Spain.
}

\begin{document}

\maketitle

\pagestyle{myheadings}

\thispagestyle{empty}

\begin{abstract}
We present a unified geometric framework for describing
both the Lagrangian and Hamiltonian formalisms of contact
autonomous mechanical systems, which is
based on the approach of the pionnering work of R. Skinner and R. Rusk. 
This framework permits to skip the second order differential equation problem, which
is obtained as a part of the constraint algorithm (for singular or regular Lagrangians),
and is specially useful to describe singular Lagrangian systems.
Some examples are also discussed to illustrate the method.
\end{abstract}

  {\bf Key words}:  Lagrangian and Hamiltonian formalisms,
contact mechanics, contact manifolds.

\bigskip

\vbox{\raggedleft AMS s.\,c.\,(2010): 
{\it Primary\/}: 37J55, 53D10, 70H03, 70H05.\\
{\it Secondary\/}: 37J05, 55R10, 70G45.}\null

\markright{{\rm M. de Le\'on} {\it et al\/},
    {\sl Unified formalism for contact systems.}}

\medskip
\setcounter{tocdepth}{2}
{
\def\baselinestretch{0.97}
\small
\def\addvspace#1{\vskip 1pt}
\parskip 0pt plus 0.1mm
\tableofcontents
}

\newpage

\section{Introduction}

In a seminal paper in 1983, R. Skinner and R. Rusk introduced a new framework for the dynamics of
first-order autonomous mechanical systems which combined
the Lagrangian and Hamiltonian formalisms \cite{SR-83}
into a single one.
The aim of this formulation was to obtain a common framework 
for both regular and singular dynamics, describing simultaneously 
the Hamiltonian and the Lagrangian formulations of the dynamics. 
Over the years, Skinner--Rusk's framework was subsequently generalized in many directions. 
So, in \cite{CMC-2002} it was extended for explicit time-dependent systems using a jet bundle language,
in \cite{GM-05} to other kinds of more general time-dependent singular differential equations, and in
\cite{BEMMR-2008,CMC-2002}
to  first-order non-autonomous dynamical systems in general.
In \cite{CLMM-2002} the Skinner--Rusk formalism was used
to study vakonomic mechanics and the comparison between the
solutions of vakonomic and nonholonomic mechanics. 
The formalism was also extended to higher-order autonomous and non-autonomous
mechanical systems \cite{art:Gracia_Pons_Roman91,
art:Gracia_Pons_Roman92,art:Prieto_Roman11,
art:Prieto_Roman12_1}, and it was also applied to control systems \cite{BEMMR-2007,art:Colombo_Martin_Zuccalli10}.
Finally, in
\cite{art:Campos_DeLeon_Martin_Vankerschaver09,LMM-2003,
ELMMR-04,PR-2015,RRS-2005,RRSV-2011,art:Vitagliano10} 
the Skinner--Rusk model was developed for first and higher-order classical field theories and,
in particular, it was used to describe different models of gravitational theories
\cite{Ca-2018,CGRS-2019,GR-2018}.

In recent years, there has been an increasing interest in the study of contact Hamiltonian and Lagrangian systems
\cite{Bravetti2017,BCT-2017,DeLeon2019,DeLeon2019b,
DeLeon2016b,GGMRR-2019b,LL-2018}.
The essential tool is contact geometry \cite{BHD-2016,BGG-2017,CNY-2013,Geiges-2008},
which has been used to describe dissipative systems 
\cite{CG-2019,Galley-2013,MPR-2018,Ra-2006}
and several other types of physical systems
in thermodynamics, quantum mechanics, circuit theory, control theory, etc.\
(see for instance,
\cite{Bravetti-2019,Goto-2016,KA-2013,RMS-2017}).
Recently, a generalization of contact geometry has been developed
to describe field theories with dissipation \cite{GGMRR-2019,GGMRR-2020}.

In the contact setting the corresponding motion equations 
are obtained using the Herglotz principle instead of the Hamilton one \cite{He-1930,Her-1985}, 
so that these dynamical systems do not enjoy conservative properties,  but dissipative ones.
The main difference between both variational principles is that, 
in the Herglotz variational principle, the action is defined 
by a non-autonomous ODE instead of an integral. 
Therefore, if we start with a Lagrangian function
$L\colon\Tan Q \times \R \longrightarrow \R$ 
such that $L = L (q^i, v^i, z)$ using bundle coordinates,
then the solutions to the dynamics obey the Herglotz equations
$$
\frac{d}{dt}\left(\frac{\partial L}{\partial \dot{q}^i}\right) - \frac{\partial L}{\partial q^i} = \frac{\partial L}{\partial \dot{q}^i} \frac{\partial L}{\partial z} \ ,
$$
where $v^i=\dot q^i$, and they are sometimes called {\sl generalized Euler-Lagrange equations}.

The contact Hamiltonian picture is obtained on the bundle 
$\Tan^*Q \times \R$ just considering the canonical contact form
$\eta=\d z-\theta_o$, where $\theta_o=p_i\,\d q^i$ 
(in bundle coordinates) is the canonical Liouville form on 
$\Tan^*Q$. So, given a Hamiltonian function 
$H\colon\Tan ^* Q\times\R\longrightarrow\R$,
we can find a unique Hamiltonian vector field satisfying the equations
$$
\inn(X_H)\d\eta=\d H-({\Reeb}(H))\eta
\quad ,\quad
            \inn(X_H)\eta=-H\ ,
$$
where ${\cal R}$ is the {\sl Reeb vector field} characterized by the conditions 
$$
\inn({\cal R})\d\eta = 0 \quad ,\quad 
\inn({\cal R})\eta = 1 \ .
$$
The integral curves of $X_H$ satisfy the contact Hamiltonian equations
$$
\frac{dq^i}{dt} = \frac{\partial H}{\partial p_i} \quad , \quad
\frac{dp_i}{dt} = -\left(\frac{\partial H}{\partial q^i} + p_i \frac{\partial H}{\partial z}\right) \quad ,
\quad \frac{dz}{dt} = p_i \frac{\partial H}{\partial p_i} - H \ .
$$
When the Lagrangian $L$ is regular (in the usual sense) we can pass from the Lagrangian to the Hamiltonian picture by means of the corresponding Legendre transformation.

The aim of this paper is to extend the Skinner-Rusk formalism to contact dynamical
systems (Section \ref{uf}), now, carefully studying the dynamical equations of
motion and the submanifold where they are consistent,
and showing how the Lagrangian and Hamiltonian descriptions
are recovered from this unified framework.

First, we define the {\sl extended unified bundle}
(also called the {\sl extended Pontryagin bundle})
${\cal W}=\Tan Q\times_Q\Tan^*Q\times\R$.
Then we consider a precontact form on ${\cal W}$, 
which is just the pull-back of the canonical contact form
on $\Tan^*Q\times\R$. 
Finally, the Hamiltonian energy is constructed from a Lagrangian 
$L\in\Cinfty(\Tan Q\times \R)$ by
$$
\H =p_i v^i-L(q^j,v^j,z)\in\Cinfty({\cal W}) \ .
$$
The rest is just to apply a constraint algorithm to this precontact Hamiltonian system. 
One of the main interest in such formulation is that the SODE condition is obtained for free. 
If the Lagrangian is regular, we obtain the usual results 
when the dynamics is projected on the Lagrangian or the Hamiltonian side. 
In the singular case, the algorithm is properly connected (also by projection)
with the corresponding Lagrangian and Hamiltonian constraint algorithms.

The paper is structured as follows. 
Section 2 is devoted to recall the main facts and results on contact Hamiltonian and Lagrangian dynamics. 
In section 3 we develop the unified formalism and explain
how the Lagrangian and Hamiltonian descriptions are recovered from it.
Finally, in Section 4, we discuss several interesting examples of regular and singular systems.

All the manifolds are real, second countable and $\Cinfty$. The maps are assumed to be $\Cinfty$. Sum over repeated indices is understood.

\section{Hamiltonian and Lagrangian formalisms of contact systems}
\label{prel}

\subsection{Contact geometry and contact Hamiltonian systems}

(See, for instance,
\cite{BCT-2017,GGMRR-2019b,Geiges-2008,LL-2018} 
for details).
			
\begin{definition}
\label{dfn-contact-manifold}
    Let $M$ be a $(2n+1)$-dimensional  manifold. 
    A \textbf{contact form} in $M$ is a differential $1$-form
    $\eta\in\df^1(M)$ such that $\eta\wedge(\d\eta)^{\wedge n}$
    is a volume form in $M$.
    Then $(M,\eta)$ is said to be a \textbf{contact manifold}.
\end{definition}

The fact that
$\eta\wedge(\d\eta)^{\wedge n}$ is a volume form 
induces a decomposition
$$
\Tan M= \ker\d\eta\oplus\ker\eta\equiv\mathcal{D}^{\rm R}\oplus\mathcal{D}^{\rm C}\ . 
$$

\begin{prop}
If $(M,\eta)$ is a contact manifold
then there exists a unique vector field $\Reeb\in\vf(M)$, 
which is called \textbf{Reeb vector field}, 
such that
\begin{equation}\label{eq-Reeb}
        \inn(\Reeb)\d\eta = 0\quad ,\quad
        \inn(\Reeb)\eta = 1.
\end{equation}
This vector field generates the distribution ${\cal D}^{\rm R}$, 
which is called the \textbf{Reeb distribution}.
\end{prop}

In addition, for every point $p\in M$, there exist a chart 
$(U; q^i, p_i, z)$, $1\leq i\leq n$, such that
$$
\eta\vert_U= \d z - p_i\,\d q^i \quad ; \quad
\Reeb\vert_U= \frac{\partial}{\partial z}\  .
$$
These are the \textsl{Darboux} or \textsl{canonical coordinates} of the contact manifold $(M,\eta)$ \cite{Go-69}.

The canonical model for contact manifolds is the manifold 
$\Tan^*Q\times\R$. 
In fact, if $z$ is the cartesian coordinate of $\R$, and 
$\theta_o\in\df^1(\Tan^*Q)$ 
and $\omega_o=-\d\theta_o\in\df^2(\Tan^*Q)$
are the canonical forms in $\Tan^*Q$,
and $\pi_1\colon \Tan^*Q \times\R \to \Tan^*Q$ 
is the canonical projection, then 
$\eta=\d z-\pi_1^*\theta_o$ is a contact form in 
$\Tan^*Q\times\R$,with $\d\eta=\pi_1^*\omega_o$,
and the Reeb vector field is
$\displaystyle\Reeb= \frac{\partial}{\partial z}$.

Given a contact manifold $(M,\eta)$, 
we have the $\Cinfty(M)$-module isomorphism
\begin{equation*}
\begin{array}{rccl}
    \flat\colon & \vf(M) & \longrightarrow & \df^1(M) \\
    & X & \longmapsto & \inn(X)\d\eta+(\inn(X)\eta)\eta
\end{array}
\end{equation*}

\begin{teor}
\label{teo-hameqs}
    If $(M,\eta)$ is a contact manifold, for every $H\in\Cinfty(M)$,
     there exists a unique vector field $X_H\in\vf(M)$ such that
    \begin{equation}
\label{hamilton-contact-eqs}
            \inn(X_H)\d\eta=\d H-({\Reeb}(H))\eta
\quad ,\quad
            \inn(X_H)\eta=-H \ .
    \end{equation}
    Then, the integral curves ${\bf c}\colon I\subset\R\to M$ of $X_H$
    are the solutions to the equations
    \begin{equation}\label{hamilton-contactc-curves-eqs}
            \inn({\bf c}')\d\eta=\left(\d H-({\Reeb}(H))\eta\right)\circ{\bf c}
\quad ,\quad
            \inn({\bf c}')\eta=-H\circ{\bf c} \ ,
    \end{equation}
    where ${\bf c}'\colon I\subset\R\to \Tan M$ is the canonical lift of the curve
    ${\bf c}$ to the tangent bundle $\Tan M$.
\end{teor}

\begin{definition}
The vector field $X_H$ is the
\textbf{contact Hamiltonian vector field} associated to $H$ and the equations 
\eqref{hamilton-contact-eqs} and \eqref{hamilton-contactc-curves-eqs}
are the \textbf{contact Hamiltonian equations} 
for this vector field and its integral curves, respectively.
The triple $(M,\eta,H)$ is said to be a \textbf{contact Hamiltonian system}.
\end{definition}

Taking Darboux coordinates $(q^i,p_i,z)$, 
the contact Hamiltonian vector field is
$$ 
X_H = \frac{\partial H}{\partial p_i}\frac{\partial}{\partial q^i} - 
\left(\frac{\partial H}{\partial q^i} + 
p_i\frac{\partial H}{\partial z}\right)\frac{\partial}{\partial p_i} + 
\left(p_i\frac{\partial H}{\partial p_i} - H\right)\frac{\partial}{\partial z}\ ; 
$$
and its integral curves ${\bf c}(t) = (q^i(t), p_i(t), z(t))$
are solutions to the dissipative Hamilton equations  
\eqref{hamilton-contactc-curves-eqs} which are
\beq
    \dot q^i = \frac{\partial H}{\partial p_i}\quad ,\quad
    \dot p_i = -\left(\frac{\partial H}{\partial q^i} + p_i\frac{\partial H}{\partial z}\right)\quad ,\quad
    \dot z = p_i\frac{\partial H}{\partial p_i} - H\ .
\label{Heqcoor}
\eeq

\begin{remark}
{\rm The contact Hamiltonian equations \eqref{hamilton-contact-eqs} are equivalent to 
\begin{equation*}
    \Lie({X_H})\eta= -(\Reeb(H)) \, \eta
    \quad ,\quad
    \inn(X_H)\eta = -H \:,
\end{equation*}
and also to
\begin{equation*}
\flat(X_H) =\d H - (\Reeb(H) + H) \eta \ .
\end{equation*}

Furthermore, equations \eqref{hamilton-contact-eqs} can be written 
without making use of the Reeb vector field $\Reeb$,
as follows:
consider the open set $U=\{p\in M;H(p)\not= 0\}$ and
the 2-form $\Omega = -H\,\d\eta + \d H\wedge\eta$ on~$U$.
A vector field $X_H\in\vf(U)$ is the contact Hamiltonian vector field if, and only if,
$$
\inn(X_H)\Omega = 0
\quad,\quad
\inn(X_H)\eta = -H\,.
$$
On the open set $U$, a path ${\bf c}\colon I\subset\R\to M$ 
is an integral curve of the 
contact Hamiltonian vector field $X_H$
if, and only if, it is a solution to
$$
\inn({\bf c}')\Omega = 0
\quad , \quad
\inn({\bf c}')\eta = - H\circ{\bf c} \:.
$$
}
\end{remark}

\begin{remark}{\rm
When some of the conditions stated in Definition 
\ref{dfn-contact-manifold} do not hold,
$\eta$ is said to be a
{\sl precontact structure} and $(M,\eta)$ is a {\sl precontact manifold}
(then the map $\flat$ is not an isomorphism)
and $(M,\eta,H)$ is called a {\sl precontact Hamiltonian system}.
Then, the Hamiltonian equations are not necessarily compatible everywhere on $M$ 
and a suitable {\sl constraint algorithm} must be implemented in order to find 
a {\sl final constraint submanifold} $P_f\hookrightarrow M$
(if it exists) where there are Hamiltonian vector fields $X_H\in\vf(M)$,
tangent to $P_f$, which are (not necessarily unique) solutions to the Hamiltonian equations on $P_f$.
Furthermore, for precontact manifolds, 
Reeb vector fields are not uniquely determined but,
if $(M,\eta,H)$ is a precontact Hamiltonian system,
the constraint algorithm and the final dynamics
are independent on the Reeb chosen.
 (See \cite{DeLeon2019} for a deeper analysis on all these topics).
}\end{remark}

\subsection{Contact Lagrangian systems}
\label{sec-conLagsys}

(See  \cite{CG-2019,CIAGLIA2018,DeLeon2019,GGMRR-2019b}
 for details).

Let $Q$ be an $n$-dimensional manifold and
the bundle $\Tan Q\times\R$ with canonical projections
$$
z\colon \Tan Q\times\R\to\R 
\quad , \quad
\tau_1\colon \Tan Q\times\R\to\Tan Q
 \quad , \quad
\tau_0\colon \Tan Q\times\R\to Q\times\R\ .
$$
Natural coordinates in $\Tan Q\times\R$ are denoted $(q^i,v^i,z)$.

As a product manifold, we can write
$\Tan(\Tan Q \times \R) =
(\Tan(\Tan Q)\times \R)\oplus_{\Tan Q\times\Real}(\Tan Q\times\Tan\R)
$,
so any operation that can act on tangent vectors to $\Tan Q$
can act on tangent vectors to $\Tan Q \times \R$.
In particular, the vertical endomorphism 
of $\Tan(\Tan Q)$ and the Liouville vector field on $\Tan Q$ yield a
\textsl{vertical endomorphism}
${\cal J} \colon \Tan (\Tan Q\times\R) \to \Tan (\Tan Q\times\R)$ and a \textsl{Liouville vector field}
$\Delta \in \vf(\Tan Q\times\R)$
(this is the Liouville vector field 
of the vector bundle structure defined by $\tau_0$).
In natural coordinates,
their local expressions are
$$
{\cal J} =
\frac{\partial}{\partial v^i} \otimes \d q^i
\,,\quad
\Delta = 
v^i\, \frac{\partial}{\partial v^i} \ .
$$

Let ${\bf c} \colon\R \rightarrow Q\times\R$ 
be a path, with ${\bf c} = (\mathbf{c}_1,\mathbf{c}_0)$.
The \textsl{prolongation} of ${\bf c}$ to $\Tan Q\times\R$ 
is the path
$$
{\bf \tilde c} = (\mathbf{c}_1',\mathbf{c}_0)
\colon 
\R \longrightarrow \Tan Q \times \R  \,,
$$
where $\mathbf{c}_1'$ is the velocity of~$\mathbf{c}_1$.
The path ${\bf \tilde c}$ is said to be \textsl{holonomic}.
A vector field  $\Gamma \in\vf(\Tan Q\times\R)$ 
is said to satisfy the \textsl{second-order condition}
(for short: it is a {\sc sode}) when all of its integral curves 
are holonomic. 
In coordinates, if 
${\bf c}(t)=(c^i(t), z(t))$, then
$$
{\bf \tilde c}(t) =
\left( c^i(t),\frac{d c^i}{d t}(t),z(t) \right) \:.
$$
and the local expression of a {\sc sode} is
$$
\Gamma= 
v^i \frac{\partial}{\partial q^i} +
f^i \frac{\partial}{\partial v^i} + 
g\,\frac{\partial}{\partial z}
\ .
$$
So, in coordinates a {\sc sode} defines a system of
differential equations of the form
$$
\frac{\d^2 q^i}{\d t^2}=f^i(q,\dot q,z) \:, \quad  
\frac{d z}{d t}=g(q,\dot q,z)  \:.
$$

A vector field $\Gamma\in\vf(\Tan Q\times\R)$ 
is a \textsc{sode} if, and only if,
${\cal J}(\Gamma)=\Delta$.

\begin{definition}
\label{lagrangean}
A \textbf{Lagrangian function} 
is a function $L\colon\Tan Q\times\R\to\R$.
\\    
The \textbf{Lagrangian energy}
associated with $L$ is the function $E_L := \Delta(L)-L\in\Cinfty(\Tan Q\times\R)$.
\\    
The \textbf{Cartan forms}
associated with $L$ are defined as
$$
\theta_L = 
{}^t{\cal J} \circ \d L \in \Omega^1(\Tan Q\times\R) 
\:,\quad
\omega_L = -\d \theta_L\in \Omega^2(\Tan Q\times\R) \ .
$$
The \textbf{contact Lagrangian form} is
$$
\eta_L=\d z-\theta_L\in\Omega^1(\Tan Q\times\R)
\:;
$$
it satisfies that $\d\eta_L=\omega_L$.
\\
The couple $(\Tan Q\times\R,L)$ is a \textbf{contact Lagrangian system}.
\end{definition}

In natural coordinates in $\Tan Q\times\R$ we have
\beann
\eta_L&=& \d z - \frac{\partial L}{\partial v^i} \,\d q^i \:,
\\
\d\eta_\L &=& 
-\frac{\partial^2L}{\partial z\partial v^i}\d z\wedge\d q^i 
-\frac{\partial^2L}{\partial q^j\partial v^i}\d q^j\wedge\d q^i 
-\frac{\partial^2L}{\partial v^j\partial v^i}\d v^j\wedge\d q^i\ ,
\eeann

Now, we define the {\sl Legendre map} 
associated with a Lagrangian $L$ as the fiber derivative of $L$,
considered as a function on the vector bundle
$\tau_0 \colon \Tan Q\times\R \to Q \times \R$;
that is, the map
${\cal F}L\colon \Tan Q\times\R\to\Tan^*Q\times\R$
given by
$$
{\cal F}L({\rm v}_q,z) = \left( \strut {\cal F}L(\cdot,z) ({\rm v}_q),z \right)
\,,
$$
where $L(\cdot,z)$ is the Lagrangian with $z$ fixed.
Its local expression in natural coordinates is
$$
  {\cal F}L^*z = z \quad\ , \quad
  {\cal F}L^*q^i = q^i \quad\  , \quad
  {\cal F}L^*p_i =\derpar {\Lag}{v^i} \ .
$$

\begin{remark}\rm
The Cartan forms can also be defined as
$\theta_L={\cal F}L^{\;*}(\pi_1^*\theta_o)$ and
$\omega_L={\cal F}L^{\;*}(\pi_1^*\omega_o)$.
\end{remark}

\begin{prop}
\label{Prop-regLag}
Given a Lagrangian $L$, then the Legendre map
${\cal F}L$ is a local diffeomorphism if, and only if,
$(\Tan Q\times\R,\eta_L)$
is a contact manifold.
\end{prop}

The conditions in the proposition mean that the Hessian matrix 
$\displaystyle (W_{ij})=\left( \frac{\partial^2L}{\partial v^i\partial v^j}\right)$ 
is everywhere nonsingular.

\begin{definition}
A Lagrangian function $L$ is said to be \textbf{regular} if the equivalent
conditions in Proposition \ref{Prop-regLag} hold.
Otherwise $L$ is called a \textbf{singular} Lagrangian.
In particular, 
$L$ is said to be \textbf{hyperregular} 
if ${\cal F}L$ is a global diffeomorphism.

A singular Lagrangian is \textbf{almost-regular} if:
(i) $P_1={\cal F}L(\Tan Q\times\R)$
is a closed submanifold of $\Tan^*Q\times\R$,
(ii) ${\cal F}L$ is a submersion onto its image,
(iii) for every ${\rm v}_q\in\Tan Q\times\R$, the fibres
 ${\cal F}L^{-1}({\cal F}L({\rm v}_q))$
 are connected submanifolds of $\Tan Q\times\R$.
\end{definition}

\begin{remark}{\rm
As a result of the preceding definitions and results,
every regular contact Lagrangian system 
has associated the contact Hamiltonian system
$(\Tan Q\times\R, \eta_L, E_L)$.
}
\end{remark}

Given a regular contact Lagrangian system 
$(\Tan Q\times\R,L)$, from \eqref{eq-Reeb} 
we have that the \textsl{Reeb vector field} 
$\Reeb_L\in\vf(\Tan Q\times\R)$ 
for this system is uniquely determined by the relations
$$
    \inn(\Reeb_L)\d\eta_L=0 \quad ,\quad
    \inn(\Reeb_L)\eta_L=1 \ ,
$$
and its local expression is
$$
\Reeb_L=\frac{\partial}{\partial z}-W^{ji}
\frac{\partial^2L}{\partial z \partial v^j}\,\frac{\partial}{\partial v^i} \ ,
$$
where $(W^{ji})$ is the inverse of the Hessian matrix,
namely 
$W^{ji} W_{ik} = \delta^j_k$.

\begin{definition}
\label{def-lageqs}
Let $(\Tan Q\times\R,L)$ be a contact Lagrangian system.
\\
The \textbf{contact Euler--Lagrange equations} for a holonomic curve
${\bf\tilde c}\colon I\subset\R \to\Tan Q\times\R$ are
\begin{equation}
\inn({\bf\tilde c}')\d\eta_L = 
\Big(\d E_L - (\Reeb_L(E_L))\eta_L\Big)\circ{\bf\tilde c} 
\quad ,\quad
\inn({\bf\tilde c}')\eta_L = - E_L\circ{\bf\tilde c} 
\ ,
\label{hec}
\end{equation}
where ${\bf\tilde c}'\colon I\subset\R\to\Tan(\Tan Q\times\R)$ denotes the
canonical lifting of ${\bf\tilde c}$ to $\Tan(\Tan Q\times\R)$.
\\
The \textbf{contact Lagrangian equations} for a vector field $X_L\in\vf(\Tan Q\times\R)$ are 
\begin{equation}
\label{eq-E-L-contact1}
    \inn(X_L)\d \eta_L=\d E_L-(\Reeb_L(E_L))\eta_L
\quad , \quad
    \inn(X_L)\eta_L=-E_L \ .
\end{equation}
A vector field which is a solution to these equations is called a
\textbf{contact Lagrangian vector field}
(it is a contact Hamiltonian vector field for the function $E_L$).
\end{definition}

\begin{remark}{\rm
In the open set $U=\{p\in M;\H(p)\not= 0\}$,
the above equations can be stated equivalently as
$$
\inn({\bf \tilde c}')\Omega_L = 0 \quad ,\quad
\inn({\bf \tilde c}')\eta_L = - E_L\circ{\bf\tilde c} \ ,
$$
and
$$
           \inn(X_L)\Omega_L = 0
\quad ,\quad
            \inn(X_L)\eta_L = -E_L\ ,
$$
where $\Omega_L=-E_L\,\d\eta_L + \d E_L\wedge\eta_L$.
}\end{remark}

In natural coordinates, for a holonomic curve
${\bf\tilde c}(t)=(q^i(t),\dot q^i(t),z(t))$,
the contact Euler-Lagrange equations \eqref{hec} are
\bea
\label{ELeqs2}
\dot z&=&L \ ,
 \\
\label{ELeqs3}
\displaystyle\frac{\partial^2L}{\partial v^j \partial v^i}\,
\ddot q^j +\displaystyle\frac{\partial^2L}{\partial q^j \partial v^i} \,\dot q^j  +
\displaystyle \frac{\partial^2L}{\partial z\partial v^i}\, \dot z
-\displaystyle\frac{\partial L}{ \partial q^i}=
\frac{d}{dt}\left(\derpar{L}{v^i}\right)-
\displaystyle\frac{\partial L}{\partial q^i}&=&
\displaystyle\frac{\partial L}{\partial z}\displaystyle\frac{\partial L}{\partial v^i} \ ;
\eea
meanwhile, for a vector field $X_L\in\vf(\Tan Q\times\R)$,
if $L$ is a regular Lagrangian, then $X_L$ is a {\sc sode}
which is called the \textsl{Euler--Lagrange vector field} 
associated with $L$ and whose integral curves are
the Euler--Lagrange equations \eqref{ELeqs2} and \eqref{ELeqs3}.
The local expression of this Euler--Lagrange vector field is
$$
X_L=L\,\frac{\partial}{\partial z}
+v^i\,\frac{\partial}{\partial q^i}
+W^{ik}
\left(
\frac{\partial L}{ \partial q^k} 
- \frac{\partial^2L}{\partial q^j \partial v^k} \,v^j
- L\frac{\partial^2L}{\partial z \partial v^k} 
+\frac{\partial L}{\partial z}
 \frac{\partial L}{\partial v^k} 
\right)
\frac{\partial}{\partial v^i} \ .
$$

\begin{remark}{\rm
If $L$ is singular, although $(\Tan Q\times\R,\eta_L)$ is not
strictly a contact manifold, but a precontact one, and hence the Reeb vector field
is not uniquely defined, it can be proved
that the Lagrangian equations \eqref{eq-E-L-contact1} 
are independent on the Reeb vector field used  (see \cite{DeLeon2019}).
Then, solutions to the Lagrangian equations
are not necessarily {\sc sode} and,
in order to obtain the Euler--Lagrange equations \eqref{ELeqs3}, 
the condition ${\cal J}(X_L)=\Delta$ must be added to the above Lagrangian equations.
Furthermore, these equations are not necessarily compatible everywhere on $\Tan Q\times\R$ 
and a suitable {\sl constraint algorithm} must be implemented in order to find 
a {\sl final constraint submanifold} $S_f\hookrightarrow\Tan Q\times\R$
(if it exists) where there are {\sc sode} vector fields $X_L\in\vf(\Tan Q\times\R)$,
tangent to $S_f$, which are (not necessarily unique) solutions to the above equations on $S_f$.
All these problems have been studied in detail in  \cite{DeLeon2019}.
}
\end{remark}

\begin{remark}{\rm
In the (hyper)regular case we have that ${\cal F}L$ is a
diffeomorphism between $(\Tan Q\times\R)$ and $(\Tan^*Q\times\R)$,
and ${\cal F}L^{\;*}\eta=\eta_L$.
Furthermore, there exists (maybe locally) a function $H\in\Cinfty(\Tan^* Q\times\R)$ 
such that ${\cal F}L^{\;*}H=E_L$; then we have the
contact Hamiltonian system $(\Tan^*Q\times\R,\eta,H)$,
for which ${\cal F}L_*{\Reeb}_\L={\Reeb}$.
Then, if $X_H\in\vf(\Tan^*Q\times\R)$ is the
contact Hamiltonian vector field associated with $H$,
we have that ${\cal F}L_*X_L=X_H$.

In the almost-regular case we have the submanifold
$j_1\colon P_1={\cal F}L(\Tan Q\times\R)\hookrightarrow\Tan^* Q\times\R$,
and ${\cal F}L^{\;*}\eta=\eta_L$.
Then there exists a function $H_1\in\Cinfty(P_1)$ 
such that ${\cal F}L^{\;*}H_1=E_L$, and we have the
precontact Hamiltonian system $(P_1,\eta_1,H_1)$,
where $\eta_1=j_1^*\eta$.
The corresponding (precontact) Hamilton equations are not necessarily compatible everywhere on $P_1$ 
and a {\sl constraint algorithm} must be implemented in order to find 
a {\sl final constraint submanifold} $P_f\hookrightarrow P_1$
(if it exists) where there are vector fields $X_{H_1}\in\vf(P_1)$,
tangent to $P_f$, which are (not necessarily unique) solutions to the above equations on $P_f$.
This algorithm and the equivalence between the Lagrangian and the Hamiltonian description
of these precontact systems are also studied in  \cite{DeLeon2019}.
}
\label{rem6}
\end{remark}


\section{Unified formalism}
\protect\label{uf}

\subsection{Unified bundle: precontact canonical structure}

For a contact dynamical system the {\sl configuration space}
is $Q\times\R$, where $Q$ is an $n$-dimensional manifold,
with coordinates $(q^i,z)$.
Then, consider the bundles $\Tan Q\times\R$ and $\Tan^* Q\times\R$ 
with canonical projections
\beann
\tau_1\colon \Tan Q\times\R\to\Tan Q & , &
\tau_0\colon \Tan Q\times\R\to Q\times\R \\
\pi_1\colon \Tan^*Q\times\R\to\Tan^*Q & , &
\pi_0\colon \Tan^*Q\times\R\to Q\times\R \ ,
\eeann
with natural coordinates $(q^i,v^i,z)$ and $(q^i,p_i,z)$ adapted to the bundle structures.
We denote by $\d z$ the volume form in $\Real$, and its pull-backs
to all the manifolds.
Let $\theta_o\in\df^1(\Tan^*Q)$ and
$\omega_o=-\d\theta_o\in\df^2(\Tan^*Q)$ be
the canonical forms of $\Tan^*Q$ whose local expressions are
$\theta_o=p_i\d q^i$ and $\omega_o=\d q^i\wedge\d p_i$;
and denote $\theta:=\pi_1^*\theta_o$ and
$\omega:=\pi_1^*\omega_o$.

\begin{definition}
We define the \textbf{extended unified bundle}
(also called the \textbf{extended Pontryagin bundle})
$$
{\cal W}=\Tan Q\times_Q\Tan^*Q\times\R \ ,
$$
which is endowed with the natural submersions
$$
 \rho_1\colon{\cal W}\to\Tan Q\times\R \ ,\
 \rho_2\colon{\cal W}\to \Tan^*Q\times\R \ ,\
\rho_0\colon{\cal W}\to Q\times\R \ ,\ 
z\colon{\cal W}\to \Real \ .
\label{project}
$$
\end{definition}

The natural coordinates in ${\cal W}$ are $(q^i,v^i,p_i,z)$.

\begin{definition}
We say that a path $\mbox{\boldmath $\sigma$}\colon\R\rightarrow{\cal W}$
is \textbf{holonomic} in ${\cal W}$ if the path 
$\rho_1\circ\mbox{\boldmath $\sigma$}\colon\R\to\Tan Q\times\R$ is holonomic.

A vector field $X \in\vf({\cal W})$ 
is said to satisfy the \textbf{second-order condition} in ${\cal W}$
(for short: it is a {\sc sode} in ${\cal W}$) when all of its integral curves 
are holonomic in ${\cal W}$. 
\end{definition}

In coordinates, a holonomic path in ${\cal W}$ is expressed as 
$$
\mbox{\boldmath $\sigma$}=
\left(\sigma_1^i(t),\frac{d\sigma_1^i}{d t}(t),\sigma_{2\,i}(t),\sigma_0(t) \right) \ ,
$$
and a {\sc sode} in ${\cal W}$ reads as
$$
X=v^i \frac{\partial}{\partial q^i}+F^i \frac{\partial}{\partial v^i}+G_i \frac{\partial}{\partial p_i}+f\,\frac{\partial}{\partial z} \ .
$$

The bundle ${\cal W}$ is endowed with the following canonical structures:

\begin{definition}
\ben
\item
The \textbf{coupling function} in ${\cal W}$ is the
map ${\cal C}\colon{\cal W}\to\R$  defined as follows: for every
$w=({\rm v}_q,{\rm p}_q,z)\in{\cal W}$, where $q\in Q$,
${\rm p}_q\in\Tan^*Q$, and ${\rm v}_q\in\Tan Q$, then
${\cal C}(w):=\langle{\rm p}_q,{\rm v}_q\rangle$.
\item
The \textbf{canonical $1$-form}
is the $\rho_0$-semibasic form
$\Theta:=\rho_2^*\,\theta\in\df^1({\cal W})$.
The \textbf{canonical $2$-form} is
$\Omega:=-\d\Theta=\rho_2^*\,\omega\in\df^2({\cal W})$.
\item
The \textbf{canonical contact $1$-form}
is the $\rho_1$-semibasic form
$\eta:=\d z-\Theta\in\df^1({\cal W})$.
Then $\d\eta=\Omega$.
\een
\label{coupling}
\end{definition}

In natural coordinates of ${\cal W}$ we have that
$$
\eta= \d z - p_i\,\d q^i \quad ,\quad
\d\eta=\d q^i\wedge\d p_i \ .
$$

\begin{definition}
Given a Lagrangian function $L\in\Cinfty(\Tan Q\times\R)$, let
$\Lag=\rho_1^*L\in\Cinfty({\cal W})$.
We define the \textbf{Hamiltonian function}
\beq
\H:={\cal C}-\Lag =p_i v^i-\Lag (q^j,v^j,z)\in\Cinfty({\cal W}) \ .
\label{Hamf}
\eeq
\end{definition}

\begin{remark}{\rm
Observe that $\eta$ is a precontact form in ${\cal W}$. 
Hence, $({\cal W},\eta)$ is a precontact manifold and
$({\cal W},\eta,\H)$ is a precontact Hamiltonian system.

As a consequence, equations \eqref{eq-Reeb} do not have a
unique solution and the Reeb vector fields are not uniquely defined.
In fact, in natural coordinates of ${\cal W}$ the general solution to \eqref{eq-Reeb} are the vector fields
$\displaystyle \Reeb=\derpar{}{z} + F^i\derpar{}{v^i}$
for arbitrary coefficients $F^i$.
Nevertheless, as we have pointed out, 
the formalism is independent on the choice of these Reeb vector fields.
In our case, as ${\cal W}=\Tan Q\times_Q\Tan^*Q\times\R$
is a trivial bundle over $\R$,
the canonical vector field $\displaystyle \derpar{}{z}$ of $\R$
can be lifted canonically to a vector field in ${\cal W}$,
which can be taken as a representative of the family of
Reeb vector fields.}
\end{remark}

\subsection{Contact dynamical equations}
\protect\label{des}

\begin{definition}
The \textbf{Lagrangian-Hamiltonian problem} associated with the contact system
$({\cal W},\eta,{\cal H})$ consists in finding the integral curves of a vector field
$X_\H\in\vf({\cal W})$ satisfying that
$\flat(X_\H) =\d\H - (\Reeb(\H) + \H)\eta$;
that is,
which is a solution to the contact Hamiltonian equations
\begin{equation}
\label{Whamilton-contact-eqs}
\inn(X_\H)\d\eta=\d\H-({\Reeb}(\H))\eta
\quad ,\quad
\inn(X_\H)\eta=-\H \ .
\end{equation}
or, what is equivalent,
$$
    \Lie({X_\H})\eta= -(\Reeb(\H)) \, \eta
    \quad ,\quad
    \inn(X_\H)\eta = -\H \ ,
$$

Then, the integral curves $\mbox{\boldmath $\sigma$}\colon I\subset\R\to{\cal W}$ of $X_\H$,
    are the solutions to the equations
    \begin{equation}\label{Whamilton-contactc-curves-eqs}
\inn(\mbox{\boldmath $\sigma$}')\d\eta=\left(\d\H-({\Reeb}(\H))\eta\right)\circ\mbox{\boldmath $\sigma$}
\quad ,\quad
 \inn(\mbox{\boldmath $\sigma$}')\eta=-\H\circ\mbox{\boldmath $\sigma$} \ .
    \end{equation}
\end{definition}

As $({\cal W},\eta,\H)$ is a precontact Hamiltonian system,
these equations are not compatible everywhere in ${\cal W}$,
and we need to implement the standard constraint algorithm
in order to find the final constraint submanifold of ${\cal W}$
(if it exists) where there are consistent solutions
to the equations. Next we detail this procedure. 

In a natural chart in ${\cal W}$, the local expression of
a vector field $X_\H\in\vf({\cal W})$ is
\beq
X_\H = f^i\derpar{}{q^i}+F^i\derpar{}{v^i}+G_i\derpar{}{p_i}+f\derpar{}{z} \ ;
\label{coorvf}
\eeq
and therefore we obtain that
\beann
\inn(X_\H)\eta&=&f-f^ip_i \ , \\
\inn(X_\H)\d\eta&=&f^i\,\d p_i-G_i\,\d q^i \ .
\eeann
Furthermore,
\beann
\d\H&=&v^i\d p_i+\left(p_i -\derpar{\Lag}{v^i}\right)\d v^i-\derpar{\Lag}{q^i}\,\d q^i-\derpar{\Lag}{z}\,\d z \ , \\
({\Reeb}(\H))\eta&=&-\derpar{\Lag}{z}(\d z-p_i\d q^i) \ .
\eeann
Then, the second equation \eqref{Whamilton-contact-eqs} gives
\beq
f=(f^i-v^i)\,p_i+\Lag \ ,
\label{zero}
\eeq
and the first equation \eqref{Whamilton-contact-eqs} leads to:
\bea
&\mbox{\rm coefficients in $\d p_i$} :& f^i=v^i \ ,
\label{one} \\
&\mbox{\rm coefficients in $\d v^i$} :& p_i=\derpar{\Lag}{v^i} \ ,
\label{two} \\
&\mbox{\rm coefficients in $\d q^i$} :& G_i=\derpar{\Lag}{q^i}
+p_i\derpar{\Lag}{z} \ ,
\label{three}
\eea
and the equalities from the coefficients in $\d z$ hold identically.
From these equations, first we have that:
\bit
\item
The equations (\ref{one}) are the holonomy conditions
(i.e., $X_\H$ is a {\sc sode}).
Thus, as it is usual, the {\sc sode} condition arises
straightforwardly from the unified formalism.
This property reflects the fact that this geometric condition in the unified formalism is stronger than in the standard Lagrangian formalism.
\item
The algebraic equations (\ref{two}) are compatibility conditions
defining a submanifold ${\cal W}_1\hookrightarrow{\cal W}$,
which is the {\sl first constraint submanifold} of the Hamiltonian precontact system $({\cal
W},\eta,\H)$, and is the graph of ${\cal F}L$; that is,
$$
{\cal W}_1=\{ ({\rm v}_q,{\cal F}L({\rm v}_q))\in{\cal W}\,\mid\,{\rm v}_q\in\Tan Q\} \ .
$$
In this way, the unified formalism includes the definition of
the Legendre map as a consequence of the constraint algorithm.
\eit
Therefore, vector fields solution to \eqref{Whamilton-contact-eqs}
are of the form
$$
X_\H=
v^i\derpar{}{q^i}+F^i\derpar{}{v^i}+
\left(\derpar{\Lag}{q^i}
+p_i\derpar{\Lag}{z}\right)\derpar{}{p_i}
+\L\derpar{}{z} 
\quad \mbox{\rm (on ${\cal W}_1$)}\ .
$$
Next, the constraint algorithm continues by demanding that
$X_\H$ must be tangent to ${\cal W}_1$,
to ensure that dynamic trajectories remain in ${\cal W}_1$. As 
$\displaystyle \xi^1_j=p_j-\derpar{\Lag}{v^j}\in\Cinfty({\cal W})$
are the constraints defining ${\cal W}_1$, this condition is
\beq
X_\H\left(p_j-\derpar{\Lag}{v^j}\right)=
-\frac{\partial^2\L}{\partial q^i\partial v^j}v^i
-\frac{\partial^2\L}{\partial v^i\partial  v^j}F^i-
\L\frac{\partial^2\L}{\partial z\partial v^j}
+\derpar{\L}{q^j}+p_j\derpar{\Lag}{z}=0
\quad \mbox{\rm (on ${\cal W}_1$)} \ .
\label{tangcond}
\eeq
At this point we have to distinguish:
\bit
\item
If $L$ is a regular Lagrangian, these equations allow us to determine all the
functions \(\displaystyle F^i=~\frac{d v^i}{d t}\);
then the solution is unique and the algorithm ends.
\item
If $L$ is singular, then these equations establish relations
among the arbitrary functions $F^i$:
some of them remain undetermined and the solutions are not unique.
Eventually, new constraints $\xi^2_\mu\in\Cinfty({\cal W})$
can appear, defining a new submanifold 
${\cal W}_2\hookrightarrow{\cal W}_1\hookrightarrow{\cal W}$
and then the algorithm continues  by demanding that
$X_\H$ must be tangent to ${\cal W}_2$, and so on
until we obtain a final constraint submanifold ${\cal W}_f$ (if it exists)
where tangent solutions $X_\H$ exist.
\eit

Now, if $\mbox{\boldmath $\sigma$}(t)=(q^i(t),v^i(t),p_i(t),z(t))$ 
is an integral curve of $X_\H$, we have that 
$\displaystyle f^i=\frac{d q^i}{d t}$,
$\displaystyle F^i=\frac{d v^i}{d t}$, 
$\displaystyle G_i=\frac{d p_i}{d t}$,
$\displaystyle f=\frac{d z}{d t}$, 
and then the equations (\ref{zero}), (\ref{one}), (\ref{two}), and (\ref{three}) 
lead to the coordinate expression of the equations
\eqref{Whamilton-contactc-curves-eqs}; in particular:
\bit
\item
From (\ref{one}), we have that $v^i=\dot q^i$;
that is, the holonomy condition.
\item
Using (\ref{one}) again, the equation (\ref{zero}) leads to
\beq
\dot z=\Lag \ ,
\label{zdot}
\eeq
which is just the equation \eqref{ELeqs2}.
\item
The equations \eqref{three} read
$$
\dot p_i=\derpar{\Lag}{q^i}+p_i\derpar{\Lag}{z}=
-\left(\derpar{\H}{q^i}+p_i\derpar{\H}{z}\right) \ ,
$$
which are the second group of Hamilton's equations \eqref{Heqcoor}.
Then, using \eqref{two} (that is, on ${\cal W}_1$),
these equations are
$$
\frac{d}{dt}\left(\derpar{L}{v^i}\right)=\derpar{\Lag}{q^i}
+\derpar{\Lag}{v^i} \derpar{\Lag}{z} \ ,
$$
which are the Euler-Lagrange equations (\ref{ELeqs3}).
The first group of Hamilton's equations \eqref{Heqcoor} arises
straightforwardly from the definition of the Hamiltonian function
\eqref{Hamf}, taking into account the holonomy condition.
\item
Using \eqref{two} (that is, on ${\cal W}_1$) and \eqref{zdot},
the tangency condition \eqref{tangcond} gives
again the contact Euler-Lagrange equations \eqref{ELeqs3}.
Observe that, if $L$ is singular, these equations could be incompatible.
\eit

\subsection{Recovering the Lagrangian and Hamiltonian formalisms and equivalence}
\label{recovering}

Next we study the equivalence of the unified formalism 
with the Lagrangian and Hamiltonian formalisms.

First, observe that, denoting by 
$\jmath_1\colon{\cal W}_1\hookrightarrow{\cal W}$
the natural embedding, we have that
$$
(\rho_1\circ\jmath_1)({\cal W}_1)=\Tan Q\times\R
\quad , \quad
(\rho_2\circ\jmath_1)({\cal W}_1)=P_1\subseteq\Tan^*Q\times\R \ .
$$
In particular $P_1$ is a submanifold of $\Tan^*Q\times\R$ when $L$ is an almost-regular Lagrangian (see Remark \ref{rem6})
and $P_1=\Tan^*Q\times\R$ when $L$ is hyperregular
(or an open set of $\Tan^*Q\times\R$ if $L$ is regular).
Furthermore, as ${\cal W}_1$ is the graph of the Legendre map
${\cal F}L$, it is diffeomorphic to $\Tan Q\times\R$,
being the restricted projection $\rho_1\circ\jmath_1$
this diffeomorphism.
In the same way, in the almost-regular case, for every submanifold
$\jmath_\alpha\colon{\cal W}_\alpha\hookrightarrow{\cal W}$
obtained by application of the constraint algorithm, we have
$$
(\rho_1\circ\jmath_\alpha)({\cal W}_\alpha)=S_\alpha\hookrightarrow\Tan Q\times\R
\quad , \quad
(\rho_2\circ\jmath_\alpha)({\cal W}_\alpha)=P_\alpha\hookrightarrow P_1\hookrightarrow\Tan^*Q\times\R \ ,
$$
and, as 
${\cal W}_\alpha\subseteq{\cal W}_1={\rm graph}\,{\cal F}L$,
then ${\cal F}L(S_\alpha)=P_\alpha$.
Finally, let $\jmath_f\colon{\cal W}_f\hookrightarrow{\cal W}$
the final constraint submanifold, and
$$
(\rho_1\circ\jmath_f)({\cal W}_\alpha)=S_f\hookrightarrow\Tan Q\times\R
\quad , \quad
(\rho_2\circ\jmath_f)({\cal W}_\alpha)=P_f\hookrightarrow P_1\hookrightarrow\Tan^*Q\times\R \ .
$$
We have the diagram
$$
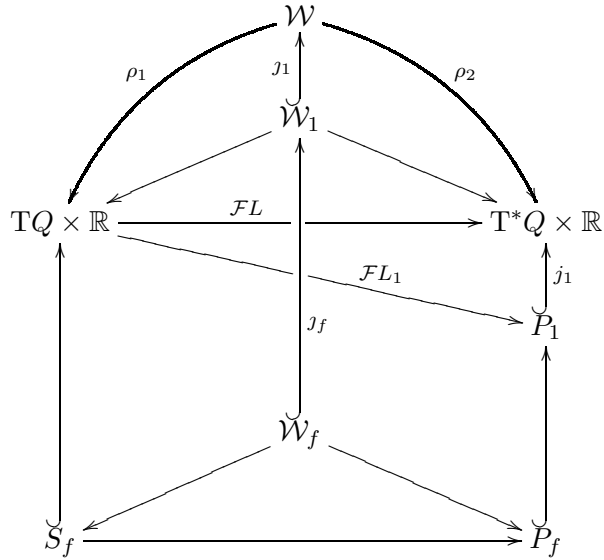
\xymatrix{
\ & \ & {\cal W} \ar@/_1.3pc/[ddll]_{\rho_1} \ar@/^1.3pc/[ddrr]^{\rho_2} & \ & \ \\
\ & \ & {\cal W}_1 \ar[dll] \ar[drr] \ar@{^{(}->}[u]^{\jmath_1} & \ & \ \\
\Tan Q\times\R \ar[rrrr]^<(0.30){{\cal F}L}|(.49){\hole}
\ar[drrrr]^<(0.60){{\cal F}L_1}|(.495){\hole} 
& \ & \ & \ & \Tan^*Q\times\R  \\
\ & \ & \ & \ & P_1 \ar@{^{(}->}[u]_{j_1} \\
\ & \ & {\cal W}_f \ar@{^{(}->}[uuu]_<(0.30){\jmath_f} \ar[dll]\ar[drr] & \ & \ \\
S_f \ar@{^{(}->}[uuu] \ar[rrrr] & \ & \ & \ & P_f \ar@{^{(}->}[uu] \\
}
$$
Every function or differential form in ${\cal W}$
and the vector fields in ${\cal W}$ tangent to ${\cal W}_1$
can be restricted to ${\cal W}_1$. 
Then, they can be translated to the Lagrangian or the Hamiltonian side 
by using that ${\cal W}_1$ is diffeomorphic to $\Tan Q\times\R$,
or projecting to the second factors of the product bundle, 
$\Tan^* Q\times\R$.
Therefore, bearing this in mind, the results and the discussion in the above section
lead to state:

\begin{teor}
Every path $\mbox{\boldmath $\sigma$}\colon I\subseteq\Real\to{\cal W}$,
taking values in ${\cal W}_1$, can be split as
$\mbox{\boldmath $\sigma$}=(\mbox{\boldmath $\sigma$}_L,\mbox{\boldmath $\sigma$}_H)$, where
$\mbox{\boldmath $\sigma$}_L=\rho_1\circ\mbox{\boldmath $\sigma$}\colon I\subseteq\Real \to\Tan Q\times\R$
and $\mbox{\boldmath $\sigma$}_H=
{\cal F}L\circ\mbox{\boldmath $\sigma$}_L\colon I\subseteq\Real\to P_1\subseteq\Tan^*Q\times\R$.

Let $\mbox{\boldmath $\sigma$}\colon\Real\to{\cal W}$,
with  ${\rm Im}\,(\mbox{\boldmath $\sigma$})\subset{\cal W}_1$,
be a path fulfilling the equations \eqref{Whamilton-contactc-curves-eqs}
(at least on the points of a submanifold
${\cal W}_f \hookrightarrow{\cal W}_1$).
Then
$\mbox{\boldmath $\sigma$}_L$ is the prolongation to
$\Tan Q\times\R$ of the projected curve
${\bf c}=\rho_0\circ\mbox{\boldmath $\sigma$}\colon\Real\to Q\times\R$ (that is, $\mbox{\boldmath $\sigma$}_L$ is
a holonomic section),
and it is a solution to the equations \eqref{hec}.
Moreover, the path 
$\mbox{\boldmath $\sigma$}_H=
{\cal F}L\circ\widetilde{\bf c}$
is a solution to the equations \eqref{hamilton-contactc-curves-eqs} (on ${\cal W}_f)$.

Conversely, for every path ${\bf c}\colon\Real\to Q\times\R$ such that
$\widetilde{\bf c}$ is a solution to the equation \eqref{hec} 
(on $S_f$), we have that the section 
$\mbox{\boldmath $\sigma$}=(\widetilde{\bf c},{\cal F}L\circ\widetilde{\bf c})$
is a solution to the equations \eqref{Whamilton-contactc-curves-eqs}.
Furthermore, ${\cal F}L\circ\widetilde{\bf c}$
is a solution to the equation \eqref{hamilton-contactc-curves-eqs}) 
 \label{mainteor1} (on $P_f$).
\end{teor}

Notice that, if $L$ is a singular Lagrangian, then these results
hold on the points of the submanifolds ${\cal W}_f$,
$S_f$ and $P_f$

As the paths $\mbox{\boldmath $\sigma$}\colon\Real\to{\cal W}$ 
solution to the equation \eqref{Whamilton-contactc-curves-eqs}
are the integral curves of holonomic vector fields $X_\H\in\vf({\cal W})$ 
solution to \eqref{Whamilton-contact-eqs},
and the paths $\mbox{\boldmath $\sigma$}_L\colon\Real\to\Tan Q\times\R$
are the integral curves of holonomic vector fields $X_L\in\vf(\Tan Q\times\R)$ 
solution to  \eqref{hec}, then an immediate corollary of the above theorem is:

\begin{teor}
Let $X_\H\in\vf({\cal W})$ be a vector field which is
solution to the equations \eqref{Whamilton-contact-eqs}
(at least on the points of a submanifold
${\cal W}_f \hookrightarrow{\cal W}_1$) and tangent to 
${\cal W}_1$ (resp. tangent to ${\cal W}_f$). Then
the vector field $X_L\in\vf(\Tan Q\times\R)$, defined by
$X_L\circ\rho_1=\Tan\rho_1\circ X_\H$,
is a holonomic vector field (tangent to $S_f$) which is a
solution to the equations \eqref{eq-E-L-contact1} (on $S_f$),
where $\H=\rho_1^*E_L$.

In addition, every holonomic vector field solution 
to the equations \eqref{eq-E-L-contact1} (on $S_f$) 
can be recovered in this way from a vector field
$X_\H\in\vf({\cal W})$ (tangent to  ${\cal W}_f$)
solution to the equations \eqref{Whamilton-contact-eqs}(on ${\cal W}_f$).
\end{teor}

The Hamiltonian formalism is recovered in a similar way,
taking into account that, now, the paths
$\mbox{\boldmath $\sigma$}_H\colon\Real\to\Tan^*Q\times\R$
are the integral curves of vector fields $X_H\in\vf(\Tan^*Q\times\R)$ 
solution to \eqref{hamilton-contact-eqs}.
So we have:

\begin{teor}
Let $X_\H\in\vf({\cal W})$ be a vector field which is
solution to the equations \eqref{Whamilton-contact-eqs}
(at least on the points of a submanifold
${\cal W}_f \hookrightarrow{\cal W}_1$) and tangent to 
${\cal W}_1$ (resp. tangent to ${\cal W}_f$). Then
the vector field $X_H\in\vf(\Tan^*Q\times\R)$, defined by
$X_H\circ\rho_2=\Tan\rho_2\circ X_\H$,
is a solution to the equations \eqref{hamilton-contact-eqs} (on $P_f$ and tangent to $P_f$),
where $\H=\rho_2^*H$.
\end{teor}

\begin{remark}{\rm
These results are the same that those obtained for the unified
formalism of non-autonomous dynamical systems.
Intrinsic proofs of the corresponding theorems can be found in
\cite{BEMMR-2008} (see also \cite{CMC-2002}).}
\end{remark}

\begin{remark}{\rm
It is important to point out that, when working with
singular Lagrangians, the equivalence between the constraint algorithms 
in the unified and in the Lagrangian formalism
only holds when the holonomy (or second-order) condition is imposed as an additional condition for the solutions in the Lagrangian case
since, unlike in the unified formalism, this condition does not hold in the Lagrangian case (see \cite{MR-92,SR-83}).}
\end{remark}


\section{Examples}
\protect\label{uex}

\subsection{General features}

In the following examples we consider some dynamical systems 
described by Lagrangians which have been modified by adding a term of dissipation 
\cite{CIAGLIA2018,GGMRR-2019b}.
So, we consider the following situation.
Let $Q$ be an $n$-dimensional differentiable manifold
and let $L=\tau_1^*L_o-\gamma z\in \Cinfty(\Tan Q\times\Real)$
be a Lagrangian, where $\gamma\in\Real$ and $L_0\in\Cinfty(\Tan Q)$ 
is a either a regular or a singular Lagrangian.
Let ${\cal W}=\Tan Q\times_Q\Tan^*Q\times\R$ be the extended unified bundle,
with local coordinates $(q^i,v^i,p_i,z)$,
and denote $\Lag=\rho_1^*L\in\Cinfty({\cal W})$
which is a regular or singular Lagrangian depending on the regularity of $L_o$
(in the singular case, we assume that it is almost-regular). Then
$$
\H=p_i v^i-L_o(q^i,v^i)+\gamma z\in\Cinfty({\cal W}) \ ,
$$
and
$$
\d\H=v^i\d p_i+\left(p_i-\derpar{L_o}{v^i}\right)\d v^i-\derpar{L_o}{q^i}\,\d q^i+\gamma\,\d z \ .
$$
Now, for a vector field $X_\H\in\vf({\cal W})$ with local expression \eqref{coorvf},
the equations \eqref{Whamilton-contact-eqs} give
\beann
f^i=v^i & , &  f=(f^i-v^i)\,p_i+\Lag=\Lag\ ,
 \\
 p_i=\derpar{L_o}{v^i} & , &
 G_i=\derpar{L_o}{q^i}-\gamma\,p_i\ .
\eeann
We have the submanifold 
${\cal W}_1={\rm graph}({\cal F}L)\hookrightarrow{\cal W}$,
and
$$
X_\H\Big\vert_{{\cal W}_1}=
v^i\derpar{}{q^i}+F^i\derpar{}{v^i}+
\left(\derpar{L_o}{q^i}-\gamma\,p_i\right)\derpar{}{p_i}
+(L_o-\gamma z)\,\derpar{}{z} \ .
$$
The tangency condition of $X_\H$ to ${\cal W}_1$ leads to
$$
X_\H\left(p_j-\derpar{L_o}{v^j}\right)=
-\frac{\partial^2L_o}{\partial q^i\partial v^j}v^i
-\frac{\partial^2L_o}{\partial v^i\partial  v^j}F^i
+\derpar{L_o}{q^j}-\gamma\,p_j=0
\quad \mbox{\rm (on ${\cal W}_1$)} \ .
$$
As remarked in Section \ref{des}, if the Lagrangian is regular,
these equations allows us to determine all the coefficients $F^i$ 
and we have a unique solution. In the singular case,
these equations establish relations
among the arbitrary functions $F^i$ and,
eventually, new constraints could appear, defining a new submanifold 
${\cal W}_2\hookrightarrow{\cal W}_1\hookrightarrow{\cal W}$.
Then, the algorithm continues until we obtain a final constraint submanifold ${\cal W}_f$ (if it exists)
where tangent solutions $X_\H$ exist.

If $\mbox{\boldmath $\sigma$}(t)=(q^i(t),v^i(t),p_i(t),z(t))$ 
is an integral curve of a solution $X_\H$ tangent to ${\cal W}_f$, 
the equations \eqref{Whamilton-contactc-curves-eqs},
on the points of ${\cal W}_f$, are in this case
$$
\dot z=L_o-\gamma z \quad , \quad
\dot q^i=v^i \quad , \quad
\dot p^i=\frac{d}{dt}\left(\derpar{L}{v^i}\right)=\derpar{L_o}{q^i}-\gamma\,p_i=\derpar{L_o}{q^i}-\gamma\,\derpar{L_o}{v^i}\ .
$$

Next we analyze three examples: one regular system 
and two singular cases, one with a unique solution and the other with multiple solutions.

\subsection{Regular example: Central force with dissipation}

Consider the system made of a particle in $\Real^3$
with mass $m$, submitted to a central potential with dissipation.
Taking $Q=\Real^3-\{(0,0,0)\}$ with local coordinates $(q^i)$,
the Lagrangian that describes the dynamics is
$$
L=\frac{1}{2}m\,v_iv^i-U(r)-\gamma z\in\Cinfty(\Tan Q\times\Real) \ ,
$$
where $v_i=g_{ij}v^j$, being $g_{ij}$ the natural extension to 
${\cal W}$ of the euclidean metric in $\Real^3$, and $r=\sqrt{q_iq^i}$.
In the extended unified bundle
${\cal W}=\Tan Q\times_Q\Tan^*Q\times\R$,
with local coordinates $(q^i,v^i,p_i,z)$,
we denote $\Lag=\rho_1^*L\in\Cinfty({\cal W})$,
which has the same coordinate expression that $L$
and is a hyperregular Lagrangian.  Then
$$
\H=p_i v^i-\frac{1}{2}m\,v_iv^i+U(r)+\gamma z\in\Cinfty({\cal W}) \ ,
$$
and
$$
\d\H=v^i\d p_i+(p_i -m\,v_i)\d v^i+\frac{U'(r)}{r}q^i\,\d q^i+\gamma\,\d z \  .
$$
Now, for a vector field $X_\H\in\vf({\cal W})$, whose local expression is \eqref{coorvf},
the equations \eqref{Whamilton-contact-eqs} give
\beann
f^i=v^i & , &  f=(f^i-v^i)\,p_i+\Lag=\Lag\ ,
 \\
 p_i=m\,v_i & , &
 G_i=-\frac{U'(r)}{r}q_i-\gamma\,p_i\ .
\eeann
Thus we have the submanifold
 ${\cal W}_1\hookrightarrow{\cal W}$
defined by
$$
{\cal W}_1=\{ (q^i,v^i,p_i,z)\in{\cal W}\,\mid\,  p_i-m\,v_i=0 \} = {\rm graph}({\cal F}L)\ .
$$
and
$$
X_\H\Big\vert_{{\cal W}_1}=
v^i\derpar{}{q^i}+F^i\derpar{}{v^i}-
\left(\gamma\,p_i+\frac{U'(r)}{r}q_i\right)\derpar{}{p_i}
+\left(\frac{1}{2}m\,v_iv^i-U(r)-\gamma z\right)\derpar{}{z} \ .
$$
Next, the tangency condition of $X_\H$ to ${\cal W}_1$ leads to
$$
X_\H(p_i-m\,v_i)=
-\gamma\,p_i-\frac{U'(r)}{r}q_i-m\,F_i=0
\ \Longleftrightarrow\ 
F^i=-\frac{1}{m}\left(\gamma\,p^i+\frac{U'(r)}{r}q^i\right)
\ \mbox{\rm (on ${\cal W}_1$)} \ ,
$$
and the algorithm finishes giving the unique solution
$$
X_\H\Big\vert_{{\cal W}_1}=
v^i\derpar{}{q^i}-\frac{1}{m}\left(\gamma\,p^i+\frac{U'(r)}{r}q^i\right)\derpar{}{v^i}-
\left(\gamma\,p_i+\frac{U'(r)}{r}q_i\right)\derpar{}{p_i}
+\Lag\derpar{}{z} \ .
$$
Therefore, if $\mbox{\boldmath $\sigma$}(t)=(q^i(t),v^i(t),p_i(t),z(t))$ 
is an integral curve of $X_\H$, the equations
\eqref{Whamilton-contactc-curves-eqs},
on the points of ${\cal W}_1$, are
$$
\dot z=\Lag \quad , \quad
\dot q^i=v^i \quad , \quad
\frac{1}{m}\,\dot p^i=\dot v^i=\ddot q^i=
-\gamma\,\dot q^i-\frac{U'(r)}{m\,r}q^i \ ;
$$
which are the Euler-Lagrange equations for the motion of
a particle in a central potential with friction.

As stated in Section \ref{recovering},
we can recover the Lagrangian and the Hamiltonian formalisms by projecting on each factor
of ${\cal W}=\Tan Q\times_Q\Tan^*Q\times\R$.
In this case, as $L$ is a hyperregular Lagrangian,
${\cal F}L\colon\Tan Q\times\R\to\Tan^*Q\times\R$ is a diffeormorphism,
and the constraint algorithm finishes in the manifold ${\cal W}_1$.
Then, in the Lagrangian formalism, we have the holonomic contact Lagrangian vector field
$$
X_L=
v^i\derpar{}{q^i}-\left(\gamma\,v_i+\frac{U'(r)}{mr}q^i\right)\derpar{}{v^i}
+\left(\frac{1}{2}m\,v_iv^i-U(r)-\gamma z\right)\derpar{}{z}\in\vf(\Tan Q\times\R) \ ,
$$
and, in the Hamiltonian formalism, we have the contact Hamiltonian vector field
$$
X_H=
\frac{p_i}{m}\derpar{}{q^i}-
\left(\gamma\,p_i+\frac{U'(r)}{r}q_i\right)\derpar{}{p_i}
+\left(\frac{p_ip^i}{2m}-U(r)-\gamma z\right)\derpar{}{z}\in\vf(\Tan^*Q\times\R)
 \ .
$$

\subsection{Singular example: Lagrange multipliers (the damped simple pendulum)}

The {\sl Lagrange multipliers method} to incorporate constraints in a system
leads to singular Lagrangians in a natural way,
since the velocities of the multipliers do not appear in the Lagrangian. 
In order to expose how to apply this formalism to system with Lagrange multipliers, we present a simple case: the pendulum under gravity with air friction.

Consider a pendulum with mass $m$ and length $l$. 
Its position in the plain of motion is given by the polar coordinates $(r,\theta)$, such that $\theta=0$ while at rest. This motion is restricted to the circumference $r=l$. The corresponding Lagrangian is
$$
L=\frac{1}{2}m(v_r^2+r^2v_\theta^2)-mgr(1-\cos\theta)+\lambda(r-l)-\gamma z\in\Cinfty(\Tan \mathbb{R}^3\times\Real) \ ,
$$
where $\lambda$ is the Lagrange multiplier and we have added a dissipative term $-\gamma z$.
It is a singular Lagrangian since the generalized velocity $v^\lambda$ does not appear in the Lagrangian.
In the extended unified bundle
${\cal W}=\Tan \mathbb{R}^3\times_{\mathbb{R}^3}\Tan^*\mathbb{R}^3\times\R$,
with local coordinates $(r,\theta,\lambda,v_r,v_\theta,v_\lambda,p_r,p_\theta,p_\lambda,z)$,
we denote $\Lag=\rho_1^*L\in\Cinfty({\cal W})$,
which has the same coordinate expression that $L$.  Then
$$
\H=p_rv_r+p_\theta v_\theta+p_\lambda v_\lambda-\frac{1}{2}m(v_r^2+r^2v_\theta^2)+mgr(1-\cos\theta)+\gamma z-\lambda(r-l)\in\Cinfty({\cal W}) \ .
$$
Now, for a vector field $X_\H\in\vf({\cal W})$, whose local expression is \eqref{coorvf},
the equations \eqref{Whamilton-contact-eqs} give
$$
\begin{array}{ccccccc}
 f=\Lag & , & & &
 \\
f_r=v_r & , & f_\theta=v_\theta & , & f_\lambda=v_\lambda & , & 
 \\
p_r=mv_r & , & p_\theta=r^2mv_\theta & , & p_\lambda=0 & , &
 \\
G_r=mrv_{\theta}^2-mg(1-\cos\theta)+\lambda-\gamma p_r 
& , & G_\theta=-mgr\sin\theta-\gamma p_\theta & , &
 G_\lambda= r-l-\gamma p_\lambda & . &
\end{array}
$$
Thus we have the submanifold
 ${\cal W}_1\hookrightarrow{\cal W}$
defined by
$$
{\cal W}_1=\{(r,\theta,\lambda,v_r,v_\theta,v_\lambda,p_r,p_\theta,p_\lambda,z)\in{\cal W}\,\mid\,   p_r=mv_r\, ,\ p_\theta=mr^2v_\theta\,, \ p_\lambda=0 \} = {\rm graph}({\cal F}L)\ ,
$$
and the vector field
\beann
X_\H\Big\vert_{{\cal W}_1}&=&
\Lag\derpar{}{z}+
v_r\derpar{}{r}+v_\lambda\derpar{}{\lambda}+
v_\theta\derpar{}{\theta}+F_r\derpar{}{v_r}+F_\theta\derpar{}{v_\theta}+F_\lambda\derpar{}{v_\lambda}+ \\ & &
(mrv_{\theta}^2-mg(1-\cos\theta)+\lambda-\gamma p_r)\derpar{}{p_r}
-(mgr\sin\theta+\gamma p_\theta)\derpar{}{p_\theta}+
(r-l-\gamma p_\lambda)\derpar{}{p_\lambda} \, .
\eeann
The tangency condition of $X_\H$ to ${\cal W}_1$ leads to
\beq
\label{eq:ex:3}
F_r=rv_\theta^2-g(1-\cos\theta)+\frac{\lambda}{m}-\gamma v_r
\ ,\ 2v_rv_\theta+rF_\theta=-g\sin\theta-\gamma r v_\theta
\ ,\ r=l
\ \mbox{\rm (on ${\cal W}_1$)} \,
\eeq
So, we recover dynamically the constraint $r=l$. 
The tangency condition to the submanifold 
${\cal W}_2$ defined by all these constraints gives
 $$
 v_r=0 \quad \mbox{\rm (on ${\cal W}_2$)} \ .
 $$ 
Imposing again the tangency condition on the
new submanifold  ${\cal W}_3$ so obtained, 
we obtain a new equation $F_r=0$,
which allows us to compute the Lagrange multiplier 
$$
\lambda=mg(1-\cos\theta)-mlv_\theta^2
  \quad \mbox{\rm (on ${\cal W}_3$)}  \ .
$$ 
This is a new constraint, and we have the submanifold 
${\cal W}_4$, where the tangency condition
leads to obtain a last constraint
$$
v_\lambda=m(3gv_\theta\sin\theta+2l\gamma v_\theta^2)
 \quad \mbox{\rm (on ${\cal W}_4$)}\ .
$$ 
Finally, the tangency condition on this constraint allows us to determine
$$
F_\lambda=mg\left(3v_\theta\cos\theta-3\frac{g}{l}\sin^2\theta-5\gamma v_\theta\sin\theta-2lg v_\theta^2\right)
 \quad \mbox{\rm (on ${\cal W}_4$)}\ ,
$$
and the algorithm finishes with the final constraint submanifold ${\cal W}_f={\cal W}_4$,
which is defined as
\beann
{\cal W}_f&=&\{(r,\theta,\lambda,v_r,v_\theta,v_\lambda,p_r,p_\theta,p_\lambda,z)\in{\cal W}\,\mid\,   p_r=mv_r\, ,\ p_\theta=mr^2v_\theta\,, \ p_\lambda=0\,,
 \\ & &
r=l\,, \ v_r=0\,,
\lambda=mg(1-\cos\theta)-mlv_\theta^2\,, \ 
v_\lambda=m(3gv_\theta\sin\theta+2l\gamma v_\theta^2) \}
\eeann
and the unique solution
\beann
X_\H\Big\vert_{{\cal W}_f}&=&
m(3gv_\theta\sin\theta+2l\gamma v_\theta^2)\derpar{}{\lambda}+
v_\theta\derpar{}{\theta}-\left(\frac{g}{l}\sin\theta+\gamma v_\theta\right)\derpar{}{v_\theta}+ \\ & &
mg\left(3v_\theta\cos\theta-3\frac{g}{l}\sin^2\theta-5\gamma v_\theta\sin\theta
-2lgv_\theta^2\right)\derpar{}{v_\lambda}
-ml(g\sin\theta+\gamma lv_\theta)\derpar{}{p_\theta}+
\\ & &
\left(\frac{1}{2}ml^2v_\theta^2-mgl(1-\cos\theta)-\gamma z\right)\derpar{}{z}
 \ .
\eeann
Observe that there are only three independent  variables: $z$, $\theta$, and $v_\theta$. 
Therefore, for an integral curve of $X_\H$, 
the second equation of \eqref{eq:ex:3}, on ${\cal W}_f$, 
gives the equation of motion for the only physical degree of freedom,
$$
\ddot\theta=-\frac gl\sin\theta-\gamma\dot\theta\ ;
$$
which is the usual equation of the damped simple pendulum.

As stated above, we can recover the Lagrangian and the Hamiltonian formalisms 
by projecting on each factor
of ${\cal W}=\Tan \mathbb{R}^3\times_{\mathbb{R}^3}\Tan^*\mathbb{R}^3\times\R$.
Thus, in the Lagrangian formalism, we have the final constraint submanifold
\beann
{S}_f=\{(r,\theta,\lambda,v_r,v_\theta,v_\lambda,z)\in{\cal W}\,&\mid&\,r=l\,, \ v_r=0\,, \lambda=mg(1-\cos\theta)-mlv_\theta^2\,, \\ & &
 \ 
v_\lambda=m(3gv_\theta\sin\theta+2l\gamma v_\theta^2) \}
\eeann
and the holonomic contact Lagrangian vector field
\beann
X_L\Big\vert_{S_f}&=&
v_\theta\derpar{}{\theta}+v_\lambda\derpar{}{\lambda}-\left(\frac{g}{l}\sin\theta+\gamma v_\theta\right)\derpar{}{v_\theta}+F_\lambda\derpar{}{v^\lambda}+
\\ & &
\left(\frac{1}{2}ml^2v_\theta^2-mgl(1-\cos\theta)-\gamma z\right)\derpar{}{z}\in\vf(\Tan\mathbb{R}^3\times\R) \ .
\eeann
Furthermore, in the Hamiltonian formalism, we have 
$$
P_f=\{ (r,\theta,\lambda,p_r,p_\theta,p_\lambda,z)\in\Tan^*\Real^3\times\Real\,\mid\,
 \ r=l\,, \ p_\lambda=0\,, \ p_r=0\,, \ \lambda=mg(1-\cos\theta)-\frac{p_\theta^2}{ml^3} \}
$$
and the contact Hamiltonian vector field
\beann
X_H\Big\vert_{P_f}&=&
\frac{p_\theta}{ml^2}\derpar{}{\theta}+\left(\frac{3g}{l^2}p_\theta\sin\theta+ \frac{2\gamma}{ml^3}p_\theta^2\right)\derpar{}{\lambda}-
\left(mlg\sin\theta+\gamma p_\theta\right)\derpar{}{p_\theta}+
\\ & &
\left(\frac{p_\theta^2}{2ml^2}-mgl(1-\cos\theta)-\gamma z\right)\derpar{}{z}
\in\vf(\Tan^*\mathbb{R}^3\times\R) \ .
\eeann

\subsection{Singular example: Cawley's Lagrangian with dissipation}

The last example is an academic model based
on a known Lagrangian introduced by R. Cawley to study 
some characteristic features of singular Lagrangians 
in Dirac's theory of constrained systems \cite{Ca-79}.

In $\Tan \mathbb{R}^3\times\Real$, 
with local coordinates $(q^i,v^i,z)$, $i=1,2,3$,
consider the Lagrangian
$$
L=v^1v^3+\frac{1}{2}q^2(q^3)^2-\gamma z \ .
$$
In the extended unified bundle
${\cal W}=\Tan \mathbb{R}^3\times_{\mathbb{R}^3}\Tan^*\mathbb{R}^3\times\R$,
with local coordinates $(q^i,v^i,p_i,z)$,
we denote $\Lag=\rho_1^*L\in\Cinfty({\cal W})$,
which has the same coordinate expression that $L$.  Then
$$
\H=p_iv^i-v^1v^3-\frac{1}{2}q^2(q^3)^2+\gamma z\in\Cinfty({\cal W}) \ .
$$
Now, for a vector field $X_\H\in\vf({\cal W})$,
with local expression \eqref{coorvf},
the equations \eqref{Whamilton-contact-eqs} give
$$
\begin{array}{ccccccccc}
&  & p_1=v_3 & , & p_2=0 & , & p_3=v_1 & , &
 \\
f=\Lag & , & f^i=v^i  & , &
G_1=-\gamma p_1 & , & 
G_2=\frac{1}{2}q^3-\gamma p_2 & , &
G_3=q^2q^3-\gamma p_3 \ . 
\end{array}
$$
Thus we have the submanifold defined by
$$
{\cal W}_1=\{(q^i,v^i,p_i,z)\in{\cal W}\,\mid\,   p_1=v_3\, ,\, p_2=0 \, , \, p_3=v_1\}= 
{\rm graph}({\cal F}L)\hookrightarrow{\cal W}\ ,
$$
and the vector fields
$$
X_\H\Big\vert_{{\cal W}_1}=
v^i\derpar{}{q^i}+F^i\derpar{}{v^i}
-\gamma p_1\derpar{}{p_1}+
\frac{1}{2}q^3\derpar{}{p_2}+
\left(q^2q^3-\gamma p_3\right)\derpar{}{p_3}+
\Lag\derpar{}{z} \, .
$$
The tangency condition of $X_\H$ to ${\cal W}_1$ 
leads to determine $F_1$ and $F_3$ and gives a new constraint,
$$
F_1=q^2q^3-\gamma p_3
\ ,\ F_3=-\gamma p_1
\ ,\ q^3=0
\quad \mbox{\rm (on ${\cal W}_1$)} \ .
$$
Imposing the tangency condition on the submanifold 
${\cal W}_2$ defined by all these constraints we obtain
 $$
 v^3=0 \quad \mbox{\rm (on ${\cal W}_2$)} \ ,
 $$
 which, bearing in mind the first constraint  $p_1=v_3$,
 implies that $p_1=0$ (on ${\cal W}_2$). 
At this point, the tangency condition holds and we
have the final constraint submanifold
$$
{\cal W}_f=\{(q^i,v^i,p_i,z)\in{\cal W}\,\mid\,   p_1=v_3=0\, ,\, p_2=0 \, , \, p_3=v_1 \, ,\,
q^3=0\}
$$
and the family of solutions
$$
X_\H\Big\vert_{{\cal W}_f}=
v^1\derpar{}{q^1}+v^2\derpar{}{q^2}
-\gamma v_1\derpar{}{v^1}+F^2\derpar{}{v^2}
-\gamma z\derpar{}{z} \ .
$$

As in the above examples,
we can recover the Lagrangian and the Hamiltonian formalisms by projecting on each factor
of ${\cal W}=\Tan \mathbb{R}^3\times_{\mathbb{R}^3}\Tan^*\mathbb{R}^3\times\R$.
Then, in the Lagrangian formalism, we have the final constraint submanifold
$$
S_f=\{ (q^i,v^i,z)\in\Tan\Real^3\times\Real\,\mid\,
 \ q^3=0\, ,\, v^3=0 \}
$$
and the holonomic contact Lagrangian vector fields
$$
X_L\Big\vert_{S_f}=
v^1\derpar{}{q^1}+v^2\derpar{}{q^2}
-\gamma v_1\derpar{}{v^1}+F^2\derpar{}{v^2}
-\gamma z\derpar{}{z}\in\vf(\Tan\mathbb{R}^3\times\R) \ .
$$
Furthermore, in the Hamiltonian formalism, we have 
$$
P_f=\{ (q^i,p_i,z)\in\Tan^*\Real^3\times\Real\,\mid\,
 \ p_1=0\, ,\, p_2=0 \, ,\, q^3=0 \}
$$
and the unique contact Hamiltonian vector field
$$
X_H\Big\vert_{P_f}=
p_3\derpar{}{q^1}+v^2\derpar{}{q^2}
-\gamma p_1\derpar{}{p_1}
-\gamma z\derpar{}{z}\in\vf(\Tan^*\mathbb{R}^3\times\R) 
 \ ,
$$
(observe that $\ker\,{\cal F}L=\left\langle \displaystyle\derpar{}{v^2}\right\rangle$).

\section{Conclusion and outlook}
\protect\label{di}

We have presented a generalized framework for describing
both 
the Lagrangian and the Hamiltonian formalism for autonomous contact dynamical systems.
The key tool consists in using the natural geometric structure 
of the manifold ${\cal W}=\Tan Q\times_Q\Tan^*Q\times\R$ 
(the {\sl unified} or {\sl Pontryagin bundle\/})
to define a precontact dynamical system,
starting from a regular or an almost-regular Lagrangian function
$L$ in $\Tan Q\times\R$.
The compatibility of the dynamical equations stated in ${\cal W}$
leads to define a submanifold ${\cal W}_1$
which is identified with the graph of the Legendre map ${\cal F}L$.
As in other situations, the contact dynamical equations in the unified formalism 
are of three classes, giving different kinds of information:

- Algebraic (not differential) equations,which, in coordinates, read
\(\displaystyle p_i=\derpar {L} {v^i}\), and determine the submanifold
${\cal W}_1$ of ${\cal W}$ where the sections solution to the
dynamical equations must take their values. 
For singular Lagrangians, the constraints defining ${\cal W}_1$,
projected by $\rho_2$, give the primary constraints of the Hamiltonian formalism; that is, The  $\rho_2$-projection 
of ${\cal W}_1$ is the image of the Legendre transformation.

- The holonomic conditions, which
in coordinates are \(\displaystyle v^i=\frac{d q^i}{d t}\).
These conditions force the dynamical trajectories to be holonomic curves. 
This property, which arise straightforwardly from the dynamical equations in the unified formalism, 
reflects the fact that, in the unified formalism, the second-order condition
is stronger than the in the standard Lagrangian formalism.

-   The contact Euler--Lagrange equations
or, equivalently, the contact Hamiltonian equations. 

As we have a precontact dynamical system, a constraint algorithm
must be implemented in order to obtain a final constraint submanifold ${\cal W}_f\hookrightarrow{\cal W}_1$
where there are consistent solutions to the contact equations 
(i.e., trajectories tangent to ${\cal W}_f$). As in the standard unified formalisms, 
if $L$ is regular, then ${\cal W}_f={\cal W}_1$.
This algorithm is related (through the natural projections) 
with the corresponding ones in the Lagrangian and the Hamiltonian sides; although in the Lagrangian case, this equivalence only holds when the second-order condition is imposed as an additional condition for the solutions.

In addition, we have also discussed several interesting examples that illustrate the behaviour of the algorithm
in the regular and singular cases.

The formalism stated here could serve as a starting point 
to set the unified formalism for $k$-contact systems in
nonconservative field theories \cite{GGMRR-2019,GGMRR-2020},
as well as in other physical systems involving contact structures.

\subsection*{Acknowledgments}

We acknowledge the financial support from the Spanish
Ministerio de Ciencia, Innovaci\'on y Universidades project
PGC2018-098265-B-C33,
the MINECO Grant MTM2016-76-072-P, 
the ICMAT Severo Ochoa projects SEV-2011-0087 and SEV-2015-0554,
and the Secretary of University and Research of the Ministry of Business and Knowledge of
the Catalan Government project 2017--SGR--932.
Manuel La\'inz wishes to thank MICINN and ICMAT for a FPI-Severo Ochoa predoctoral contract PRE2018-083203.
Manuel de Le\' on and Manuel La\'inz would also like to acknowledge the hospitality of the Department 
of Mathematics at the Universitat Polit\`ecnica de Catalunya, during their stay.


\addcontentsline{toc}{subsection}{\bf References}
\itemsep 0pt plus 1pt

\end{document}